\documentclass[aps,prl,preprint,superscriptaddress,notitlepage]{revtex4-2}

\usepackage{graphicx}
\usepackage{dcolumn}
\usepackage{bm}
\usepackage[version=4]{mhchem}
\usepackage{blindtext}
\usepackage{upgreek}
 \usepackage{setspace}

\begin{document}

	\title{Thin film TaAs: developing a platform for Weyl semimetal devices}
	\author{Jocienne N. Nelson}
	\affiliation{National Renewable Energy Laboratory, Golden, Colorado 80401, USA}
	\author{Anthony D. Rice}
	\affiliation{National Renewable Energy Laboratory, Golden, Colorado 80401, USA}
	\author{Rafa{\l} Kurleto}
	\affiliation{Dept of Physics and Center for Experiments on Quantum Materials, University of Colorado, Boulder, CO 80309}
 	\author{Amanda Shackelford}
 	\affiliation{Dept of Physics and Center for Experiments on Quantum Materials, University of Colorado, Boulder, CO 80309}
 	\author{Zachary Sierzega}
  	\affiliation{Dept of Physics and Center for Experiments on Quantum Materials, University of Colorado, Boulder, CO 80309}
  	\author{Peipei Hao}
	\affiliation{Dept of Physics and Center for Experiments on Quantum Materials, University of Colorado, Boulder, CO 80309}
	\author{Bryan Berggren}
	\affiliation{Dept of Physics and Center for Experiments on Quantum Materials, University of Colorado, Boulder, CO 80309}	
	\author{Chun-Sheng Jiang}
	\affiliation{National Renewable Energy Laboratory, Golden, Colorado 80401, USA}
	\author{Andrew G. Norman}
	\affiliation{National Renewable Energy Laboratory, Golden, Colorado 80401, USA}
 	 \author{Megan E. Holtz}
	\affiliation{Colorado School of Mines, Golden, Colorado 80401, USA} 
	\author{John S. Mangum}
	\affiliation{National Renewable Energy Laboratory, Golden, Colorado 80401, USA}
	\author{Ian A. Leahy}
	\affiliation{National Renewable Energy Laboratory, Golden, Colorado 80401, USA}
	\author{Karen N. Heinselman}
	\affiliation{National Renewable Energy Laboratory, Golden, Colorado 80401, USA}
	\author{Herv\'e Ness}
 	\affiliation{Department of Physics, Faculty of Natural and Mathematical Sciences, King’s College London, Strand, London WC2R 2LS, United Kingdom}
 	\author{Mark Van Schilfgaarde}
	\affiliation{National Renewable Energy Laboratory, Golden, Colorado 80401, USA}
    \author{Daniel S. Dessau}
	\affiliation{Dept of Physics and Center for Experiments on Quantum Materials, University of Colorado, Boulder, CO 80309}	
	\author{Kirstin Alberi}
	\email[To whom all correspondence should be addressed: ]{Kirstin.Alberi@nrel.gov}	
	\affiliation{National Renewable Energy Laboratory, Golden, Colorado 80401, USA}

	\begin{abstract}
		MX monopnictide compounds (M=Nb,Ta, X = As,P) are prototypical three-dimensional Weyl semimetals (WSMs) that have been shown in bulk single crystal form to have potential for a wide variety of novel devices due to topologically protected band structures and high mobilities. However, very little is known about thin film synthesis, which is essential to enable device applications. We synthesize TaAs(001) epilayers by molecular beam epitaxy on GaAs(001) and provide an experimental phase diagram illustrating conditions for single phase, single-crystal-like growth. We investigate the relationship between nanoscale defects and electronic structure, using angle-resolved photoemission spectroscopy, Kelvin probe microscopy and transmission electron microscopy. Our results provide a roadmap and platform for developing 3D WSMs for device applications.
	\end{abstract}
	
	\maketitle
	
    \section{Introduction}
    Three-dimensional topological semimetals (TSMs) host extraordinary properties stemming from their bandstructure, such as extremely high mobility \cite{LiangNmat2015}, conductivity \cite{KumarNatComm2019} and magnetoresistance \cite{KumarNatComm2017}. Weyl semimetals (WSMs), a sub-category of TSMs, exhibit singly degenerate linear band crossings \cite{YanAnnRevcondensmater2017} that also support the chiral anomaly. Bulk crystal synthesis was central to the discovery of TSMs. Advances in thin film synthesis are now crucial to exploiting these properties in a wide variety of applications, including photodetectors\cite{RiceAFM2022}, photovoltaics\cite{OsterhoudtNM2019}, optical switches\cite{ZhuNatComm2017}, spintronic devices\cite{SunPRL2016}, thermoelectric devices \cite{WangAFM2017}, catalysts \cite{Rajamathi2017} and quantum computing \cite{Kharzeev2019}.  
    
    In this paper we synthesize bulk-like thin films of TaAs, which was the first WSM to be discovered in 2015 \cite{XuScience2015,LvPRX2015,yang_weyl_2015} and is well studied in bulk crystal form. Bulk TaAs hosts extremely high mobility electrons ($\mu_e\approx1.8\times10^5$ cm$^2$V$^{-1}$s$^{-1}$ at 10 K) and large magnetoresistances (80,000\% at 9 T) \cite{HuangPRX2015}. However, very little work has been done to establish controlled epitaxial growth of TaAs and related isostructural MX materials. Synthesis of these compounds is particularly challenging due to the lack of lattice-matched substrates. TaP and NbP have been grown on metallic Ta and Nb layers\cite{bedoya-pinto_realization_2020}, but electrical shorting is unacceptable for devices. Growth on GaAs (001) substrates has only produced either polycrystalline \cite{Yanez2022} or ultrathin ($<$10 nm) films \cite{Sadowski2022}. This previous work clearly indicates that there is strong interest in developing WSM films and a need for a materials science road map for controllably synthesizing them in any thickness range and integrating them with semiconductors.

	We report the synthesis of epitaxial single-crystal-like, single phase films of TaAs on GaAs(001) by molecular beam epitaxy (MBE) ranging from 10--200 nm thick. We introduce an experimental growth phase diagram and comment on defects induced by epitaxial growth along with strategies to mitigate them. We also present the first magnetotransport and angle-resolved photoemission spectroscopy (ARPES) measurements of bulk-like thick films (200 nm) to evaluate the impact of defects on electronic behavior. These measurements reveal key challenges in thin film 3D WSM synthesis. We end with a discussion of strategies for developing WSMs into practical device platforms.

	\begin{figure*} 
		\includegraphics[width=1\linewidth]{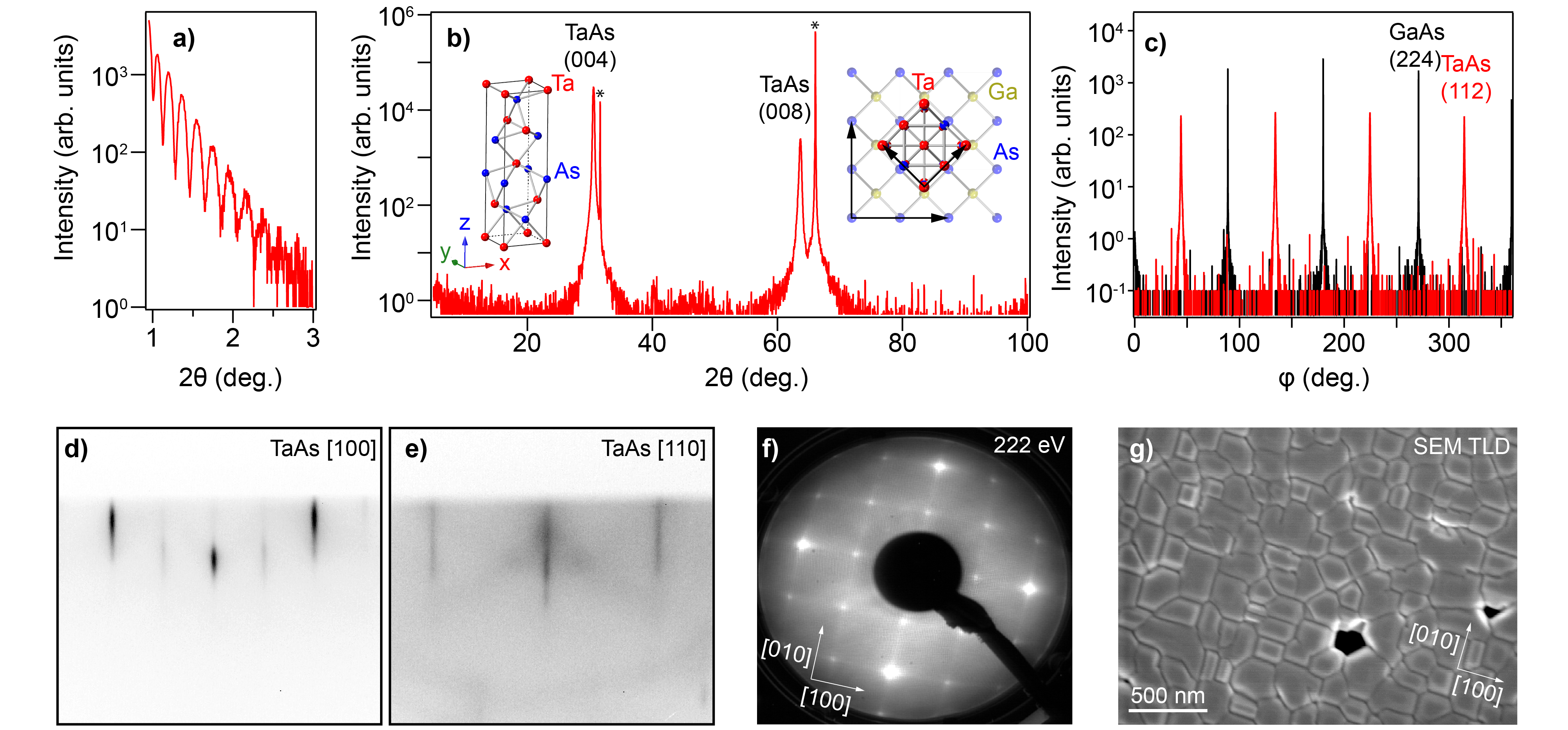}
		\caption{\label{fig:Fig1} a) X-ray reflectivity measurement of a 60 nm TaAs(001) film showing low angle Kiessig fringes. b) X-ray $\theta-2\theta$ scan of a 60 nm TaAs(001)/GaAs(001) clearly demonstrating that the sample is single phase, * indicate the substrate peaks. The left inset shows the tetragonal crystal structure of TaAs, space group No. 109, right inset is a schematic of the orientation of TaAs/GaAs showing that the TaAs is rotated 45$^\circ$ relative to the GaAs with a large 16\% mismatch between the TaAs a,b = 3.43 \AA \space and a the lattice parameter of GaAs a$/\sqrt{2}$. c) X-ray $\phi$ scan of the GaAs(224) peak (black) and the TaAs(112) peak (red). d-e) Reflection high energy electron diffraction (RHEED) images along the d) [100] and e) [110] azimuths of TaAs. The streaky pattern and change in spacing of streaks when rotating the sample in plane is consistent with a single crystal. f) Low energy electron diffraction (LEED) image measured at 222 eV. g)   Scanning electron microscopy through-lens detector (TLD) secondary electron image demonstrating grain morphology. Void regions where TaAs is missing appear as black regions.}
	\end{figure*} 
	\section{Results}
	
	{\bf Epitaxial relationship between TaAs and GaAs.} We report single-crystal-like, single phase growth of TaAs films directly on GaAs(001). X-ray reflectivity shows Kiessig fringes indicating smooth interfaces (fig.\ref{fig:Fig1}a). The $2\theta$ scan (fig.\ref{fig:Fig1}b) demonstrates that the films are single phase and (001) oriented. A $\phi$ scan of off-axis peaks (fig.\ref{fig:Fig1}c) reveals that the films are rotated by approximately 45$\pm 5^\circ$ in plane relative to GaAs. The rotation angle varies from sample to sample and is reflective of a weak epitaxial relationship. Rotation is driven by the large difference between the a,b lattice parameters of TaAs (3.43 \AA) and the lattice parameter of GaAs (5.65 \AA) and reduces the mismatch to 16\%, though this is still large enough that the TaAs films relax immediately. Reflection high-energy electron diffraction (RHEED) images measured along the (fig.\ref{fig:Fig1}d) [100] and (fig.\ref{fig:Fig1}e) [110] azimuths of TaAs are streaky, indicating two-dimensional (2D) smooth film growth. The difference in spacing of the streaks along the two azimuths is consistent with a single crystal sample. This is further suggested by sharp well defined spots observed in low energy electron diffraction (LEED) images (fig.\ref{fig:Fig1}f).
	
	There are clear streaks in the measured LEED pattern along the TaAs[100] and [010] directions. This is due to a loss of microscopic ordering which disrupts the periodicity of the crystal structure \cite{QinJACS2014}. To investigate the origin, we performed scanning electron microscopy (SEM) measurements (fig.\ref{fig:Fig1}g). There is a striking rectangular morphology reminiscent of polycrystalline material with grain boundaries. The alignment of the rectangular grains, where the main edges are along [100] and [010], causes the streaks in the LEED images.  These features change slightly from film to film but are present in all samples measured with varied growth conditions, including miscut, As species (As$_2$ or As$_4$), and thickness.  In contrast to the SEM, macroscopic diffraction probes are consistent with a single crystal. Thus our TaAs/GaAs films are `single-crystal-like' -- defined by single crystal grains that are well aligned with low-angle grain boundaries \cite{GaoACSAMI2016} and is distinct from polycrystalline material where grains are randomly oriented.  
	
	\begin{figure*} 
		\includegraphics[width=1\linewidth]{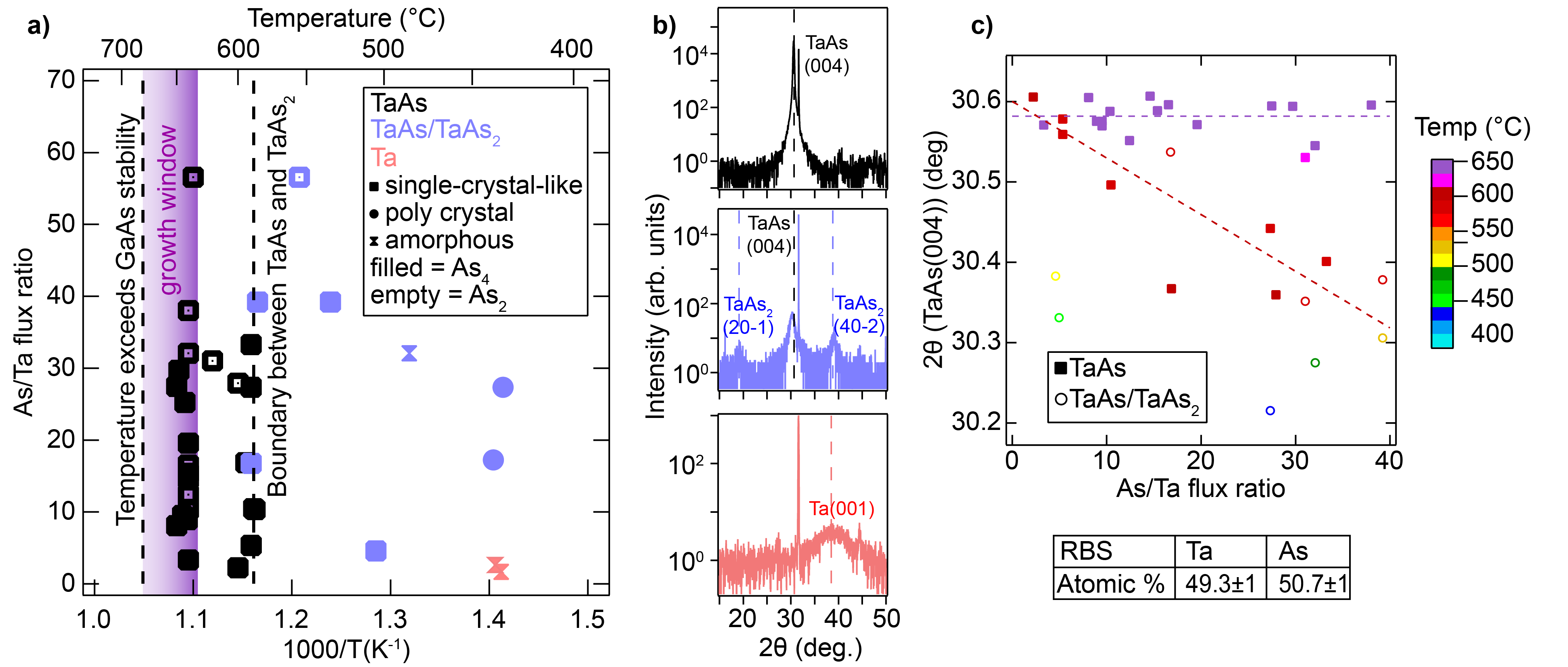}
		\caption{\label{fig:Fig2} a) Experimental phase diagram showing stability region of TaAs depending on growth temperature and As/Ta ratio. Colors indicate the phase detected by X-ray diffraction $\theta-2\theta$ scans, shown in fig.\ref{fig:Fig2}b: 1) single-crystal-like TaAs (black), 2) mixed phase TaAs$_2$/TaAs (blue), and 3) metallic Ta (red). The symbols represent the RHEED pattern, classified by single-crystal-like and streaky (squares), polycrystalline rings (circles) or amorphous (triangles). Filled symbols indicate growth with As$_4$ while empty symbols indicate growth with As$_2$. b) Representative X-ray $\theta-2\theta$ scans showing a typical single phase TaAs film (black), a mixed phase film (blue) and Ta metal (red). RHEED and XRD sample characterization is further discussed in supplementary note 1. c) TaAs(004) peak location as a function of As/Ta ratio. The color scale indicates the growth temperature, single phase films are shown as solid squares while those that display TaAs$_2$ peaks appear as open circles, and dashed lines are fits to the 640--650 $^\circ$C samples (purple) and 590--600 $^\circ$C samples (red). The table shows results from Rutherford Backscattering Spectrometry (RBS) measurement of a 200 nm TaAs film grown at 650$^\circ$C.}
	\end{figure*}

	{\bf Experimental growth phase diagram.} We now discuss growth conditions required to realize single phase TaAs films and present an experimental growth phase diagram. This is a key result of this study and may be used as a guide in the future synthesis of TaAs and for understanding defects induced by epitaxial growth. To find the region of stability for TaAs we varied the substrate temperature from 400--650 $^\circ$C and the As/Ta ratio from 1-60. The resulting film compositions are displayed in fig.\ref{fig:Fig2}a. It is apparent from this phase diagram that substrate temperature is the most important factor in stabilizing TaAs. Low temperatures (400--500 $^\circ$C) result in amorphous or polycrystalline mixed phase films (TaAs/TaAs$_2$), suggesting insufficient adatom mobility. Films grown at intermediate temperatures between 500--590 $^\circ$C also display mixed phase, but two dimensional growth indicated by streaky RHEED. Finally, at temperatures $>$590$^\circ$C, single phase TaAs is stabilized. Temperatures greater than 650 $^\circ$C were not explored in this study due to the required As overpressures necessary to prevent GaAs decomposition. Higher growth temperatures should be studied in the future using substrates capable of operating under those conditions.
	
	The stability of single-crystal-like, single phase TaAs over a large range in the As/Ta flux ratio indicates that this growth is adsorption controlled. The growth rate is limited by the Ta flux, while excess As is not incorporated into the film. This is similar to GaAs epitaxy, where stoichiometric growth may be achieved over a wide range of As/Ga flux ratios, and is consistent with other metal monopnictide systems \cite{MBE_monopnictides}. Changes in growth temperature are likely achieving two things: 1) thermodynamically shifting which phase is the most energetically stable, and 2) modifying the effective As/Ta ratio as more As desorbs at higher temperatures. To understand where the true adsorption controlled growth window lies, it is essential to identify the bounds of conditions in temperature and flux ratio where the resulting films are unchanged. It is apparent in fig.\ref{fig:Fig2}b that the TaAs(004) peak is shifted to lower $2\theta$ in the mixed phase film, likely due to strain from TaAs$_2$ inclusions which change the c axis lattice parameter. To detect changes in the TaAs structure, we track the TaAs(004) peak location as a function of As/Ta ratio (fig.\ref{fig:Fig2}c). In films grown at 640--650 $^\circ$C the peak location is constant. In contrast, lowering the growth temperature results in increasing peak shift even in samples without obvious TaAs$_2$ peaks in X-ray diffraction (XRD). Thus true adsorption controlled growth occurs at 640--650 $^\circ$C making it an optimal temperature range to synthesize TaAs. The As species (As$_2$ vs As$_4$) and the substrate miscut do not appear to impact growth. Rutherford backscattering spectrometry (RBS) measurements demonstrate that films grown in the optimized window are stoichiometric to within the uncertainty. The rest of this paper focuses on films grown at 640--650 $^\circ$C.

	\begin{figure} 
		\includegraphics[width=0.55\linewidth]{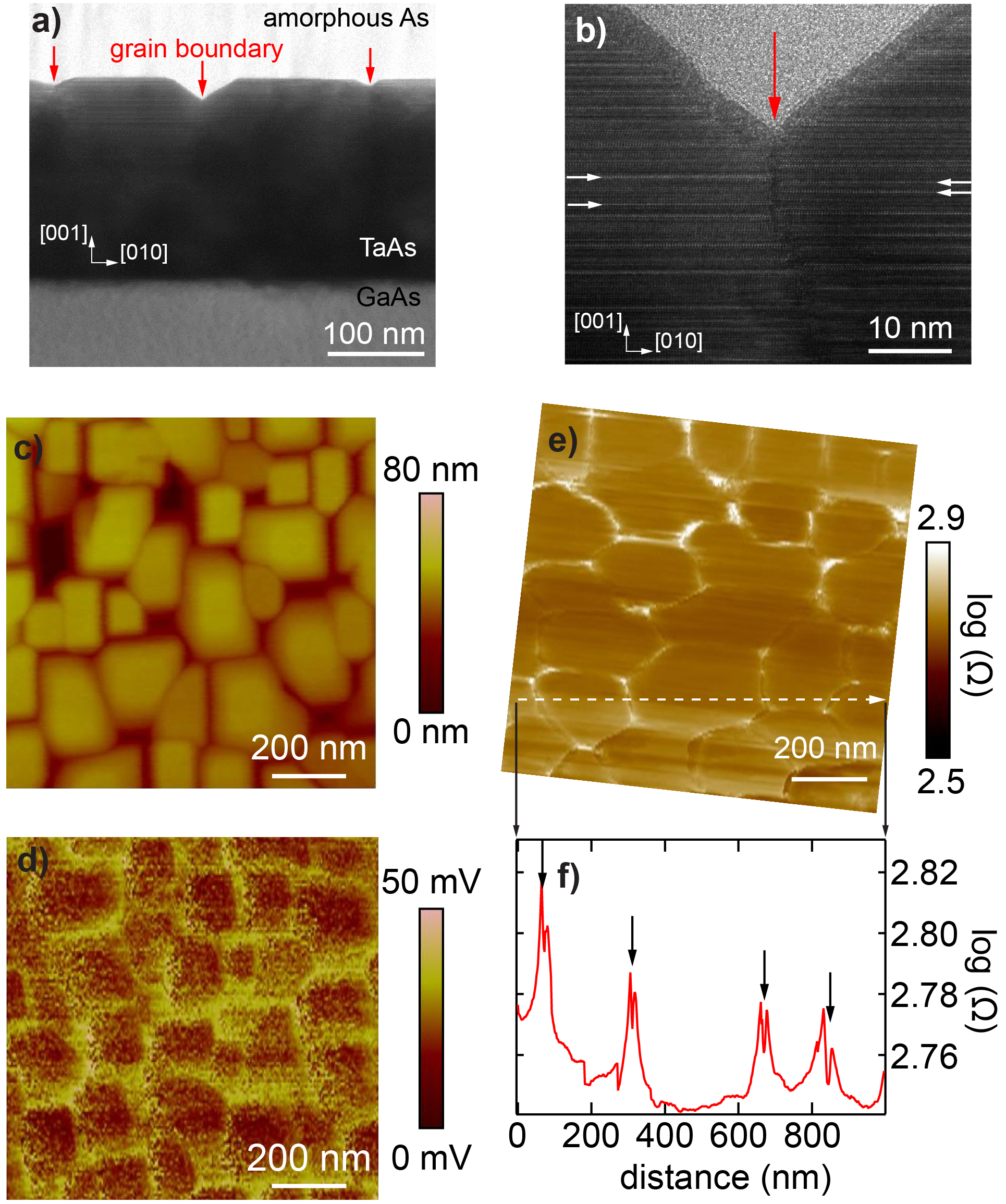}
		\caption{\label{fig:Fig3} Microscopy measurements of TaAs and defects on a 200 nm film. a) bright field transmission electron microscopy (TEM) image of TaAs/GaAs(001) with an amorphous As capping layer showing a large field of view. b) zoomed in high resolution TEM image displaying atomic resolution of the region around a grain boundary near the surface of the film. Red arrows indicate grooves at grain boundaries, while white arrows point to stacking faults, which are visible as bright horizontal lines in the TEM image. Scanning probe microscopy images c-e), c) high resolution atomic force microscopy image, d) kelvin probe force microscopy image, e) scanning spreading resistance microscopy image, the line cut below (f) is taken along the white dashed line. Each image was taken from a different region of the sample but with similar fields of view. }
	\end{figure} 	
	
    {\bf Nanoscale morphology, defects, and electronic properties.} We now focus on the defects induced by the large 16\% lattice mismatch between TaAs and GaAs. Bright field transmission electron microscopy (TEM) with contrast sensitive to defects (fig.\ref{fig:Fig3}a) shows notch-like surface grooves (red arrows) are spaced every $\sim$100--200 nm, consistent with the rectangular grains in fig.\ref{fig:Fig1}g). A high resolution TEM image (fig.\ref{fig:Fig3}b) shows atomic resolution at a grain boundary. A high density of stacking faults appears as bright horizontal lines. These are also observed in high mobility single crystals \cite{BesaraPRB2016}. The grain boundaries persist down to the GaAs interface, demonstrating that they are a result of different nucleation sites merging early during growth. 
    
 	To investigate the impact of grain boundaries on transport we perform high resolution scanning probe experiments. Atomic force microscopy (AFM) (fig.\ref{fig:Fig3}c) shows 10--20 nm deep grooves between the rectangular grains. Kelvin probe force microscopy (KPFM) (fig.\ref{fig:Fig3} d) shows a 10--20 mV surface potential increase at the boundary regions, suggesting that the boundary is positively charged. The narrowness of the boundary could mean that the potential barriers are higher and sharper than they appear in the image. Scanning spreading resistance microscopy (SSRM) (fig.\ref{fig:Fig3} e) demonstrates that the grain interior has a lower resistance than the grain boundaries. The apparent resistance changes with bias voltage polarity between the probe and sample, which shows that the grains are p-type. This is consistent with room temperature low field Hall measurements where we find a large hole concentration of $\sim1\times10^{20}-8\times10^{20}$ cm$^{-3}$. While low field Hall measurements overestimate carrier concentrations by implicitly assuming a single band model, the order of magnitude is correct. This is much larger than in bulk single crystals, where hole and electron densities are typically 10$^{18}$cm$^{-3}$ \cite{SankarJPCM2018}. A line cut (white dashed line) taken along the SSRM image (fig.\ref{fig:Fig3}f) shows an increased resistance at the grain boundaries with a dip in resistance at the center of the boundary. This behavior may be due to a depletion region around the grain boundaries, causing a reduction in hole carriers and increased resistance. Eventual crossover to n-type behavior in the middle of the boundaries then causes a dip in resistance. This very large depletion from $\sim10^{20}$ cm$^{-3}$ holes is possible because of the density of states of WSMs which becomes vanishingly small at the Weyl points.

	\begin{figure*} 
	\includegraphics[width=1\linewidth]{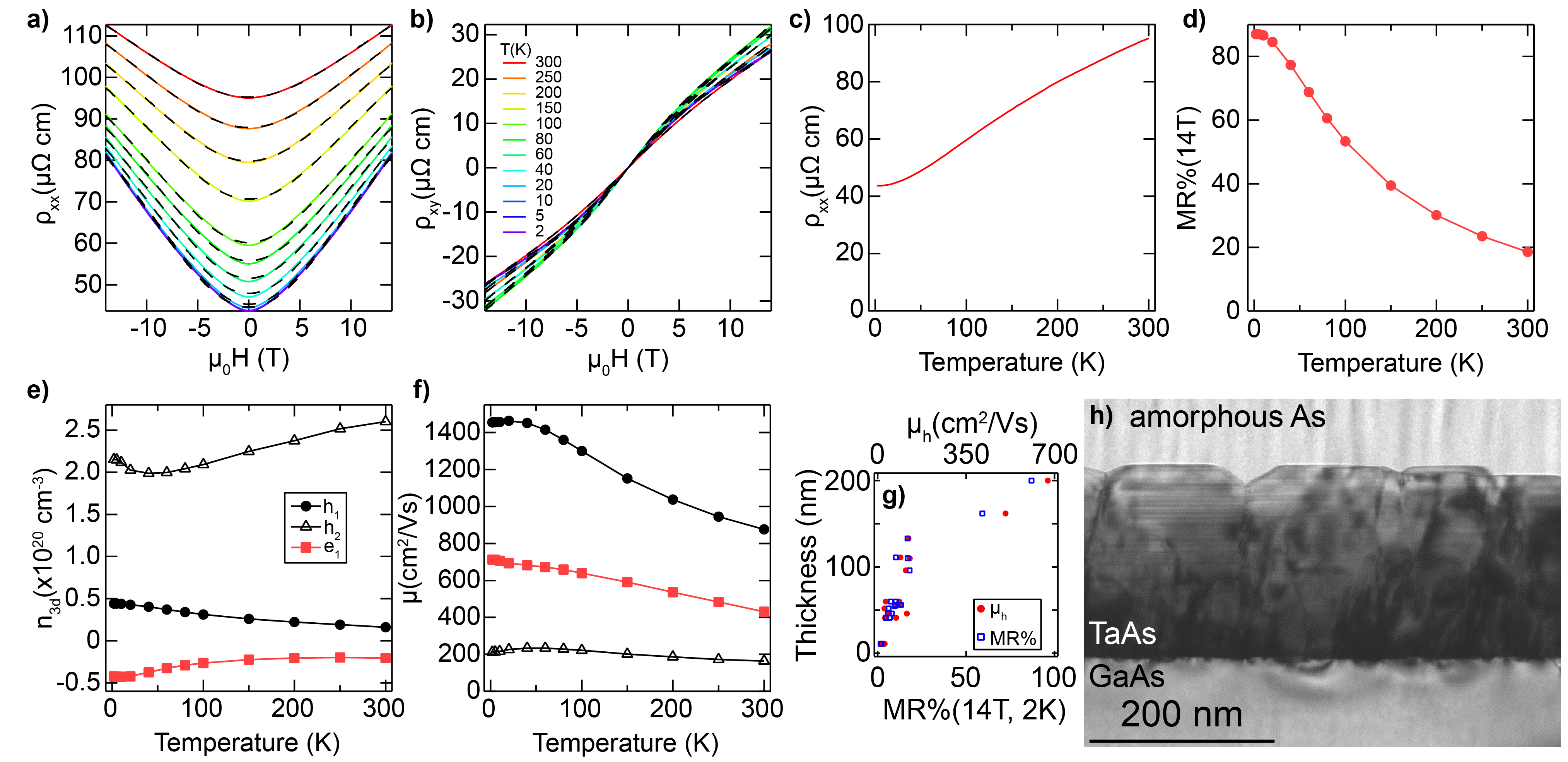}
	\caption{\label{fig:Fig4} a) $\rho_{xx}$ and b) $\rho_{xy}$ as a function of applied magnetic field  H $\parallel\hat{z}$, current I $\parallel\hat{x}$ for a 200 nm TaAs film at temperatures from 300--2 K. Black dashed lines indicate fits to a three band model. Temperature dependence of the c) zero field resistivity and d) MR\% at 14 T. e) n$_{3D}$ and f) mobility as a function of temperature extracted from the three band fit for the two hole-like carriers and one electron like carrier. g) Hall mobility at 2 K (red), and magnetoresistance \% at 14 T, 2K (blue) as a function of sample thickness. h) A diffraction contrast TEM image. Threading dislocations appear as dark streaks and are very dense near the interface with GaAs and decrease with increased layer thickness.}
\end{figure*} 
\begin{figure*} 
	\includegraphics[width=1\linewidth]{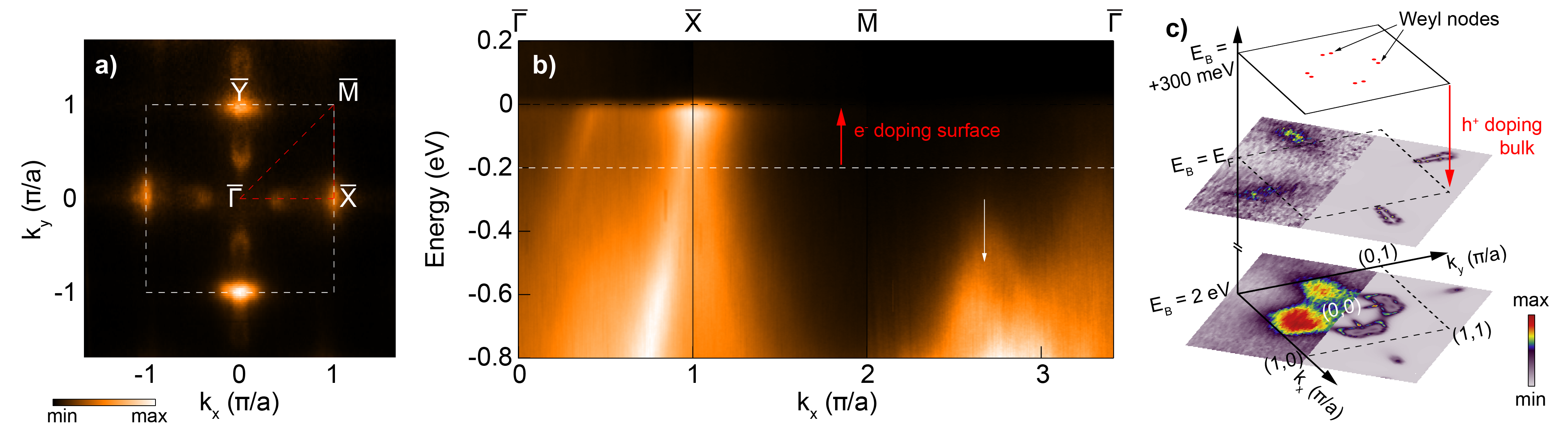}
	\caption{\label{fig:Fig5} {\bf ARPES spectra} a) VUV-ARPES Fermi surface map collected at 10 K with linear horizontal polarization showing characteristic surface states of TaAs (h$\nu=90$ eV). b) Dispersion from $\overline{\Gamma}-\overline{X}-\overline{M}-\overline{\Gamma}$ (red dashed lines in panel a) demonstrating that the Fermi level is shifted upwards by $\sim$0.2 eV relative to the surface states of bulk single crystals (h$\nu =$ 63, 90 eV). A horizontal dashed line schematically shows the estimated location of the Fermi level in bulk crystals at $\sim -0.2$ eV. c) SX-ARPES (h$\nu = 650$ eV) constant energy maps at E$_B$=E$_F$, E$_B$=2 eV on the left compared to QSGW+SOC constant energy maps at E$_B$=E$_F$-300 meV and E$_B$=-2.3 eV on the right. A schematic constant energy map at E$_B$=+300 meV hosts one set of Weyl nodes.}
\end{figure*}

{\bf Electrical properties} Magnetotransport measurements probing the longitudinal ($\rho_{xx}$) and Hall ($\rho_{xy}$) resistivity of a 200 nm thick TaAs Hall bar are shown in figure \ref{fig:Fig4} a,b. The magnetoresistance does not saturate within the field range studied, as is typical for a topological semimetal. The slope of $\rho_{xy}(H)$ is positive due to domination of hole type carriers. Non-linearity of $\rho_{xy}(H)$ indicates that multiple charge carrier types contribute to transport, as expected from the complex band structure of TaAs. Zero field resistivity measurements (fig.\ref{fig:Fig4}c) show a moderate decrease with decreasing temperature, as is typical of a semimetal. Magnetoresistance \%
(MR\%=100\% $\times \frac{(\rho_{xx}(\mu_0 H)-\rho_{xx}(\mu_0 H=0))}{\rho_{xx}(\mu_0 H=0))}$) at 14 T (fig.\ref{fig:Fig4}d) increases from 20\% at 300 K to over 80\% at 2 K. Dashed black lines in fig.\ref{fig:Fig4}a,b show a fit to a standard three band model (supplementary note 1). Extracted carrier concentration and mobility as a function of temperature are shown in figure \ref{fig:Fig4} e,f. The carriers include a high mobility electron (e$_1$) and hole (h$_1$), whose mobility increases significantly with lowered temperature. This likely drives the increase in MR\%, since mobility and MR\% are known to be linked in TSMs \cite{SinghJPM2020}. An additional hole-like carrier (h$_2$) has a large carrier concentration but much lower mobility that does not change significantly with temperature. 

We find no dependence of mobility or MR\% on the As to Ta flux ratio, miscut, or As species used. However, there is a clear dependence on the thickness of the sample (fig.\ref{fig:Fig4} g). We show the low field Hall mobility $\mu_H$ measured at $\pm$0.5 T and 2 K, and MR\% at 14 T and 2 K, for a range of TaAs samples grown at 640--650$^\circ$ C. Both $\mu_h$ and MR\% increase moderately from 10--150 nm and then increase more significantly between 150--200 nm. The $\mu_H$ decrease in ultrathin films could be related to gapping of the bulk states which is unexplored in TaAs but is known to lead to reduced mobility in related compound Cd$_3$As$_2$ below 70 nm \cite{GoyalAPLM2018}. However, it does not explain the increase in $\mu_H$ above 150 nm. To understand the microscopic mechanism we perform diffraction contrast TEM (fig.\ref{fig:Fig4}h). Strain fields which appear as arcs in the GaAs layer are likely caused by misfit dislocations due to the large lattice mismatch. We are not able to discern individual misfit dislocations. Instead, they collectively form a dark background near the GaAs/TaAs interface in this image. It is possible to distinguish threading dislocations, which appear as dark vertical streaks that are highly dense near the interface and decrease in concentration towards the surface of the film due to defect annihilation. This suggests that the decrease in extended defects drives the improvement in $\mu_H$ and MR\% in thicker films. 

{\bf Electronic structure} We have performed both vacuum ultra violet angle-resolved photoemission spectroscopy (VUV-ARPES) that is sensitive to the topological surface states and soft X-ray ARPES (SX-ARPES) that probes bulk states \cite{LVnatrevphys2019}. The VUV-ARPES Fermi surface map (fig.\ref{fig:Fig5}a) and dispersion (b) is qualitatively similar to studies of bulk single crystals\cite{XuScience2015, LvPRX2015, yang_weyl_2015}, suggesting that the surface electronic structure is preserved in thin films. However, the bands are rigidly shifted moving the Fermi level up by +0.2$\pm0.04$ eV consistent with electron doping (supplementary note 2) \cite{XuScience2015, LvPRX2015, yang_weyl_2015, BelopolskiPRL2016}. In contrast SX-ARPES displays a Fermi level shift of -0.3 eV consistent with hole doping. This is apparent in the Fermi surface at k$_z$=13.1 \AA$^{-1}$ (fig.\ref{fig:Fig5}c) which is point like in bulk material but in films consists of four lobe-like features. To understand the discrepancy we calculate the bulk band structure using the first principles Questaal package \cite{questaal:web} (supplementary note 3). We estimate that the large hole concentration measured in films would lead to a downward shift of the Fermi level by 150--400 meV. We find good agreement with maps at multiple binding energies by taking this shift into account. The combination of VUV and SX-ARPES demonstrates significant carrier depletion at the surface not present in bulk single crystals. It is consistent with the depletion at grain boundaries (fig.\ref{fig:Fig3} e) suggesting a universal behavior of defects at boundaries and surface of our films.

\section{Discussion}

We now discuss the implications of our results on the effort to develop monopnictide WSM films for device applications. We highlight key challenges to overcome along with strategies for the field to implement. 

An unintentional large hole concentration, two orders of magnitude larger than in single crystals, shifts the Fermi level away from the Weyl nodes. This reduces the impact of Weyl physics on physical properties and must be solved to develop technologies which rely on accessing the Weyl node. It appears to be generic to all studies of MX thin films. An even larger hole concentration of $>10^{22}$ cm$^{-3}$ was observed in NbP and TaP films, attributed to $>2\%$ pnictogen excess \cite{bedoya-pinto_realization_2020}. Hole concentrations of $10^{21}$-$10^{22}$ cm$^{-3}$ were observed in polycrystalline TaAs films \cite{Yanez2022}. While our films are stoichiometric, as determined by RBS, we cannot rule out the presence of point defects that modify the stoichiometry at a $<$at.\% level and are known to influence free carrier concentration. Previous experimental work on defects in TaAs has been limited to the impact of several at. \% Ta deficiency in bulk single crystals accommodated through stacking faults and point defects \cite{BesaraPRB2016, BuckeridgePRB2016}. While Ta vacancies are predicted to be the most energetically favorable under all conditions\cite{BuckeridgePRB2016}, they cause electron doping, and thus are likely not present in films. The next most likely defect to form under the As-rich growth conditions used here are As$_{\mathrm{Ta}}$ anti-site defects, which are a precursor to the formation of TaAs$_2$ and are predicted to cause hole doping \cite{ZHANGJcms2019,UllahCJP2022}. This is consistent with the larger $\sim10^{21}$ cm$^{-3}$ hole concentration we observe in samples with TaAs$_2$ inclusions (grown at 600$^\circ$C and below). Future work should include developing strategies to control point defects. In MBE, varying elemental flux ratios is one possible method. However, we varied the As/Ta ratio for growths at 640--650$^\circ$C from 1-60 without any clear impact on carrier concentration. This could indicate either that: 1) point defects are not responsible for the large hole concentration or 2) it is necessary to vary conditions more widely. This could include further increasing growth temperature which would require utilizing a different substrate due to the risk of GaAs decomposition. We have not explored Ta rich growth conditions as As vacancies are also predicted to cause hole doping \cite{BuckeridgePRB2016} and As deficiency would likely narrow the growth window. Alternatively, strategies such as substitutional electron doping to shift the chemical potential towards the Weyl nodes could be studied. However, it would not remove the underlying large defect density. 

Extended defects severely limit mobility and magnetoresistance particularly in films $<$150 nm. Achieving high quality ultrathin TaAs films is a key target. This is particularly important for microelectronics applications that rely on surface states, such as proposed WSM electrical interconnects\cite{KumarArxiv2022}. In TaAs and isostructural MX compounds, the lattice parameters are not well matched to commercially available substrates. This causes strain to be relaxed in the form of extended defects, such as misfit dislocations, threading dislocations, and stacking faults, as seen here. Interface engineering should be explored, such as using a substrate with a different lattice parameter or by including buffer layers. Our single-crystal-like samples contain aligned grains with grain boundaries which act as potential barriers and cause carrier depletion. Furthermore, grain boundaries can open backscattering channels for topological surface states \cite{LanzilloPRA2022}. Thus, they may be particularly detrimental to electronic transport in thin samples, which is surface state dominated. Most industrially-relevant growth processes produce polycrystalline material. Therefore, future work should include varying grain size and type, possibly by growing on different substrates, to understand and mitigate grain impact on properties. It will be important to disentangle the impact of extended defects we observe here such as the relative impact of stacking faults versus misfit dislocations. Furthermore, it is unclear if low angle grain boundaries in our single-crystal-like samples are as detrimental to transport as high angle boundaries in polycrystalline material. 

\section{Conclusions}

We have demonstrated synthesis of thick single-crystal-like films of TaAs/GaAs(001). The experimental phase diagram illustrates how to stabilize single phase material and will enable reproducible synthesis of TaAs and related compounds. We find that the large lattice mismatch between GaAs and TaAs leads to a local grain-like morphology with aligned grains resulting from different nucleation sites that macroscopically appear as single crystal. We have demonstrated relatively high mobility ($\sim1000$ cm$^2$V$^{-1}$s$^{-1}$) and MR\% (80 \%). Our ARPES spectra are consistent with bulk single crystals. This occurs despite a high density of defects which result in large unintentional hole doping ($\sim$10$^{20}$ cm$^{-3}$) and significant potential barriers at the boundaries. Our demonstration of films which may easily be fabricated into device structures now enables a large range of applied and fundamental studies in TaAs. The utility of TSMs will depend on understanding the impact of epitaxial growth induced defects and creating mitigation strategies. Our synthesis study provides guidance to develop these strategies and highlights specific challenges that must be overcome.

\section{Methods}

\textbf{Thin film growth and characterization}
Epitaxial films were grown on GaAs(001) substrates using a Omicron EVO25 MBE system. GaAs wafers with a 4 $^\circ$ miscut towards (111)A (4A) were used primarily. On axis, 2A, and 6A substrates were also tested but not found to impact resulting growths. The GaAs surface was prepared by a high temperature anneal at 620 $^\circ$C for 20 minutes under As$_2$ flux to remove the oxide followed by homoepitaxial growth of 500--1000 nm of GaAs at 585$^\circ$C. 
Ta and As were codeposited at a range of substrate temperatures and As/Ta flux ratios described in the main text. The As pressure at the sample position is measured using a beam flux monitor and then converted into an effective flux depending on the species. Ta flux is inferred via a combination of the emission current and sample growth rate. Optimized conditions for TaAs were found to be 640-650$^\circ$C and did not depend on the As flux or As species in the range of conditions studied. The Ga was evaporated using an effusion cell, As$_2$ and As$_4$ were evaporated from a cracker source and Ta was evaporated using a MBE Komponeten EBVV vertical electron beam evaporator. The substrate temperature was measured using a KSA BandiT temperature monitor. Film growth was monitored using a KSA 400 Reflection high energy electron diffraction (RHEED) system. The crystal structure of films were characterized using x-ray diffraction with Cu K$\alpha$. Film thickness was either measured via X-ray reflectivity thin film fringes or estimated based on the growth rate of samples grown under similar conditions for samples that were too thick to display fringes. 

\textbf{RBS}
Rutherford backscattering spectrometry (RBS) was performed using a 3 MeV He++ beam in a 168$^\circ$ backscattering geometry, using a model 3S-MR10 RBS system from National Electrostatics Corporation. The sample was rocked to find an orientation that minimized channeling effects, then a measurement was performed until the total integrated charge delivered to the sample was 80 $\mu$C. The film composition was determined through fitting the data using the RUMP analysis software\cite{BARRADAS20081338}, with Ta and As as the sole constituents in the film, and the Ga:As ratio set to 1:1 in the substrate, since Ga and As show up too close together in energy, relative to the $\sim$25 keV FWHM of the detector, to distinguish between them.

\textbf{SEM}: SEM micrographs were acquired using a secondary electron detector in a through-lens configuration on a FEI Nova NanoSEM 630 operating at 3 keV accelerating voltage and 80 pA beam current at a working distance of 4.5 mm. In the main text we show a through-lens detector (TLD) secondary electron image sensitive to film topography.

\textbf{TEM} Cross-sectional specimens for TEM analysis were prepared using focused ion beam liftout, on a ThermoFisher FEI Helios NanoLab 600i, with a final ion-beam cleaning at 2 kV to minimize surface amorphization. Ga+ ion FIB damage was subsequently removed in a Fischione Model 1040 NanoMill using $<$ 1 kV Ar+ ions with the sample cooled using a liquid nitrogen cold stage. The samples were then examined in a FEI Tecnai F20 UltraTwin field emitting gun (FEG) scanning transmission electron microscope (STEM) operated at 200 kV.

\textbf{Scanning probe measurements}
Atomic force microscopy (AFM) surface morphology and AFM-based nano-electrical imaging were conducted using BrukerNano Icon and D3100 AFMs with Nanoscope V controller. The surface morphology was imaged by tapping mode with a supersharp probe (Systems for Research, MSS-FESPA) that has a cylindrical shape of diameter d=30 nm, length l=400 nm, and tip radius r=2-3 nm, to ensure the accurate AFM imaging not being compromised by the tip-sample geometry convolution. The surface potential images were taken by a home-made mode Kelvin probe force microscopy (KPFM)\cite{NonnenmacherAPL1991} using the second harmonic oscillation of the probe cantilever (Nanosenser, PPP-EFM), to enhance the voltage sensitivity to $\sim$10 mV and spatial resolution of $\sim$30 nm \cite{KikukawaAPL1995, JiangJAP2012, JiangAPL2015}. The first harmonic oscillation was used for surface morphology imaging simultaneously with the KPFM potential imaging. Scanning spreading resistance microscopy (SSRM) resistance images were taken based on the contact mode of AFM and using a logarithm-scale amplifier with a wide current range of 10$^{-15}-10^{-3}$ A. A large force ($\sim$1 $\mu$N) was applied to the diamond-coated probe (Bruker DDESP) and a large bias voltage (4 V) was applied between the sample and probe, to minimize the probe-sample contact resistance that ensures the measured resistance being dominated by the sample’s local spreading resistance right beneath the probe with a spatial resolution less than $\sim$50 nm \cite{EybenJVST2010, JiangIEEE2022}. 

\textbf{Magnetotransport} Room temperature Hall measurements were performed in a van der Pauw geometry with excitation voltages of 5 mV. Variable temperature measurements were performed on Hall bars using a Physical Property Measurement System from Quantum Design. Hall bars were fabricated with standard photolithography, ion milling and electroplated Au contacts. 

\textbf{ARPES and LEED} Angle-resolved photoemission spectroscopy measurements in vacuum ultraviolet regime (VUV-ARPES) were performed at the Merlin beamline of Advanced Light Source (Berkeley, California). 

A protective As capping layer (thickness ~200 nm) was thermally removed from TaAs films on GaAs(001) substrates prior to ARPES measurements by heating the sample to 450$^\circ$ for 30 minutes. The quality of the surface was confirmed by observing LEED patterns at room temperature. All LEED images were recorded using Specs ErLEED setup equipped with a CCD camera. The electron beam current was varied from 39 eV and 275 eV.

VUV-ARPES spectra were collected with Scienta R8000 photoelectron energy analyzer at temperature equal to 10 K and base pressure of 10$^{-10}$ mbar and a typical energy resolution of 30 meV. Systematic photon energy scans were performed (hv = 40 – 65 eV and 80 – 120 eV) to exclude contribution of bulk bands to spectra. 

Bulk sensitive ARPES studies in the soft X-ray regime were performed at ADRESS beamline of Swiss Light Source\cite{Strocov:bf5029,Strocov:ve5018}. Spectra were collected at photon energies between 440 and 825 eV using Phoibos 225 (SPECS) analyzer. Circularly polarized radiation was used, because this maximizes the intensity in the geometry of the experimental setup. We reached energy resolution of 89 meV at hv=440 eV and 130 meV at hv=650 eV. Identical decapping procedure was applied as described previously, but without LEED.

\textbf{QSGW} First principles electronic structure calculations have been performed using the Questaal package \cite{questaal:web}.
Questaal is an all-electron method, with an augmented wave basis consisting of partial waves inside augmentation
spheres based on the linear muffin-tin orbital technique  \cite{Pashov:2020}.

Calculations have been done at different levels of density functional theory (DFT), in the local-density (LDA) and generalized gradient approximations (GGA), and beyond DFT.  We considered a primitive unit cell for bulk TaAs consisting of 4 atoms (2 atoms of Ta and 2 of As) and have performed calculations for different lattice parameters for TaAs. Calculation results are presented for lattice parameters $a=b=3.4368$ \AA, $c=11.6442$ \AA\ \cite{MURRAYJLCM1976}. LDA and GGA calculations confirm previously reported band-structures \cite{WengPRX2015,XuScience2015,LvNatPhys2015,LvPRX2015,yang_weyl_2015,SunPRB2015,huangNatComm2015}.  Going beyond DFT, we have also performed Quasiparticle Self-consistent \emph{GW} (QS\emph{GW}) calculations\cite{Faleev04}.  QS\emph{GW} provides an effective way to implement the \emph{GW} approximation without relying on a lower level theory as a starting point. This form of \emph{GW} significantly reduces discrepancies with experimental data and makes what discrepancies that remain much more uniform\cite{mark06qsgw}.

For TaAs, QS\emph{GW} yields Fermi surfaces similar to LDA.  Band dispersions are also similar, though they are slightly wider in QS\emph{GW}.  Many-body effects can also be included with \emph{GW}.  Spectral functions for interacting quasi-particles (obtained via a frequency- and \emph{k}-space dependent self-energy) permits us to extract the spectral functions that can be directly compared with ARPES.  The two-dimensional color maps (at constant energy) for the spectral functions shown in the main text are calculated in the presence of SOC.  

\section{Acknowledgements}
This work was authored in part by the National Renewable Energy Laboratory, operated by Alliance for Sustainable Energy, LLC, for the U.S. Department of Energy (DOE) under Contract No. DE-AC36-08GO28308. 

Funding to perform TaAs epitaxy and basic structural and electrical characterization (MR, SEM and XRD) were supported through an NREL Director's Fellowship. Funding for the investigation of the electronic structure and disorder induced by doping and defects (TEM, RBS, scanning probe measurements, ARPES, LEED, QSGW and DFT) were supported by the U.S. Department of Energy, Office of Science, Basic Energy Sciences, Division of Materials Sciences and Engineering, Physical Behavior of Materials Program. We thank Dr. Bill McMahon for helpful discussions about film synthesis. 

We are grateful to Jonathan Denlinger from Advanced Light Source, and to Fatima Alarab, Vladimir Strocov and Procopious Constantinou from Swiss Light Source for assistance during ARPES experiments. This research used resources of the Advanced Light Source, which is a DOE Office of Science User Facility under contract no. DE-AC02-05CH11231. We acknowledge the Paul Scherrer Institut, Villigen, Switzerland for provision of synchrotron radiation beamtime at beamline ADRESS of the SLS. We would like to thank Peipei Hao and Bryan Berggren from the University of Colorado at Boulder for their help in ARPES measurements.

The views expressed in the article do not necessarily represent the views of the DOE or the U.S. Government. The U.S. Government retains and the publisher, by accepting the article for publication, acknowledges that the U.S. Government retains a nonexclusive, paid-up, irrevocable, worldwide license to publish or reproduce the published form of this work, or allow others to do so, for U.S. Government purposes.

* Kirstin.Alberi@nrel.gov


%

\renewcommand{\thefigure}{S\arabic{figure}}
\setcounter{figure}{0}

\section{Supplementary Note 1: Film growth and characterization:}
We utilize {\it in situ} reflection high-energy electron diffraction (RHEED) to monitor the film growth and provide information used to construct the experimental phase diagram. During initial growth we observe well separated GaAs and TaAs streaks indicating that the TaAs films relax immediately and are not strained to the substrate. We observe a wide variety of RHEED patterns depending on growth conditions. Single-crystal-like TaAs always coexists with streaky RHEED patterns indicating two dimensional film growth and epitaxial orientation to the substrate. Metallic Ta always appears amorphous in RHEED (lack of any visible signal). Mixed phase samples display a variety of RHEED patterns including streaky RHEED, polycrystalline rings, and amorphous RHEED. In these samples the TaAs portion of the crystal is oriented epitaxially relative to the substrate with the [001] axis out of plane and a 45$^\circ$ rotation in plane relative to GaAs in the same manner as single phase TaAs. We determine this orientation using x-ray diffraction. The TaAs$_2$ portion of the crystal is oriented with the [20-1] axis out of plane and X-ray diffraction $\phi$ scans suggest a 90$^\circ$ twin domain in plane. TaAs$_2$ is known to display extreme magnetoresistance of $>$10$^5$\% and, to the best of our knowledge, has not previously been synthesized in thin film form \cite{WangPRB2016}.

To analyze the magnetotransport data in the main text we fit $\rho_{xx}$ and $\rho_{xy}$ to a standard three band model:

\begin{eqnarray}
	\sigma_{xx}(H) =\displaystyle\sum_{i=1} ^{3} \frac{|n_i|e\mu_i}{1+(\mu_iH)^2},\\
	\sigma_{xy}(H) = \displaystyle\sum_{i=1} ^{3}\frac{|n_i|e\mu_i^2H}{1+(\mu_iH)^2},\\
	\rho_{xx} =  \frac{\sigma_{xx}}{\sigma_{xx}^2+\sigma_{xy}^2},\\
	\rho_{xy} = \frac{\sigma_{xy}}{\sigma_{xx}^2+\sigma_{xy}^2},
\end{eqnarray}
where $\sigma_{xx}$ and $\sigma_{xy}$ are components of the conductivity tensor, $i$ indicates the carrier, $n_i$ is carrier density, $e$ is the charge of an electron, $\mu_i$ is mobility and $H$ is field. We find a significantly improved fit with three rather than two carriers.

\section{Supplementary Note 2: Angle-resolved photoemission spectroscopy:}

\begin{figure*} 
	\includegraphics[width=1\linewidth]{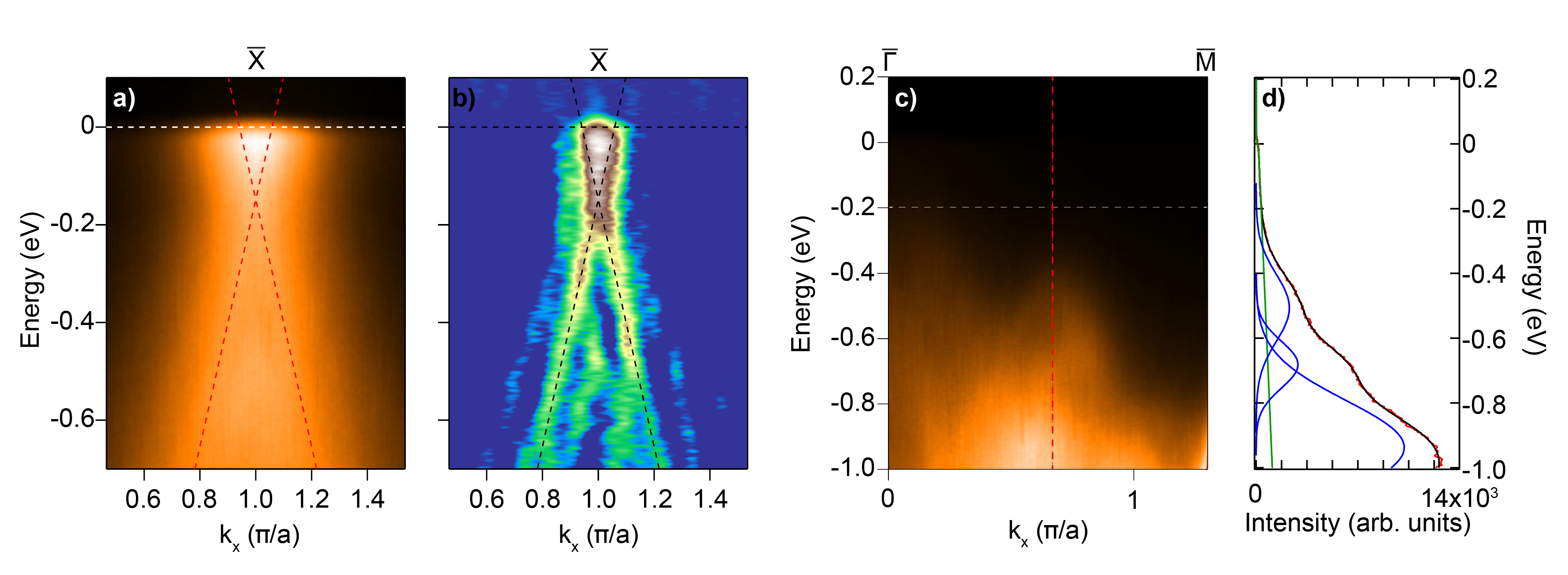}
	\caption{\label{fig:vuvARPES} a) dispersion along $\Gamma-X-\Gamma$. b) second derivative of same dispersion showing an electron-like band at Fermi level. c) dispersion along $\Gamma-M$. d) EDC along red dashed line in panel c with fit to a Fermi-Dirac distribution plus three Gaussian peaks.}
\end{figure*} 

In the VUV-ARPES Fermi surface map shown in the main text (figure 5a) characteristic spoon-like features are visible between $\overline{\Gamma}-\overline{X}$ and $\overline{\Gamma}-\overline{Y}$. Previous reports have demonstrated that these features consist of outer Fermi arc states that terminate at the bulk projection of the Weyl nodes and inner closed trivial contours\cite{LvPRX2015}. Bulk crystals display a difference between pockets centered at $\overline{X}$ and $\overline{Y}$, indicating $C4$ symmetry breaking. The lack of any discernible symmetry breaking here is likely due to the beam spot averaging over many grains in the single-crystal-like film (fig.3) which are rotated arbitrarily 90$^\circ$ from each other.   

In the main text we state that the surface states measured with VUV-ARPES display a rigid band shift causing the Fermi level to move +0.2$\pm0.04$ eV relative to reports of bulk single crystals \cite{XuScience2015, LvPRX2015, yang_weyl_2015, BelopolskiPRL2016} in a manner consistent with electron doping. Here we provide supporting analysis. The difference between films and bulk crystals can be most clearly seen in the pocket centered at $\overline{X}(\overline{Y})$. Figure \ref{fig:vuvARPES}a shows the dispersion through $\overline{X}$ along the direction $\overline{\Gamma}-\overline{X}-\overline{\Gamma}$ which clearly shows a electron-like dispersion at the Fermi level. The dispersion can be seen more clearly in the second derivative taken along the k$_x$ direction (fig. \ref{fig:vuvARPES}b). In \ref{fig:vuvARPES}a-b dashed lines are a guide to the eye showing that the hole-like linear dispersion terminates below the Fermi level. In bulk crystals this dispersion is clearly hole-like as it crosses the Fermi level. Published calculations for example Lv et. al \cite{LvPRX2015} show a Van Hove singularity with a transition to electron like bands above the Fermi level. In our films the Van Hove singularity is pushed below the Fermi level. 

It is also clear that the hole-like pockets are pushed further below the Fermi level in a manner consistent with electron doping. For example we examine the pocket between $\overline{M}-\overline{\Gamma}$, indicated by white arrow in fig.5b and shown in fig. \ref{fig:vuvARPES}c. To extract the band location we fit the energy distribution curve taken at the k$_x$ corresponding to the band maxima (along the red dashed line) to a Fermi-Dirac distribution plus three Gaussian peaks. We find that the maxima of the hole-like band is at $0.49\pm0.016$ eV. In contrast this band is approximately maxima at 0.3 eV in bulk crystals \cite{yang_weyl_2015, LvPRX2015} leading us to estimate the Fermi level shift of +0.2$\pm0.04$ eV.

\section{Supplementary Note 3: QGSW}

We have performed first principles electronic structure calculations by using the Questaal package
\cite{Pashov:2020,questaal:web}.  Calculations have been at different levels of Density Functional Theory (DFT) and
beyond DFT using the primitive unit cell for bulk TaAs consisting of 4 atoms (2 atoms of Ta and 2 of As). 

We considered different values of the lattice parameters for TaAs which can be found in the literature
\cite{WengPRX2015,XuScience2015,LvNatPhys2015, LvPRX2015,yang_weyl_2015,SunPRB2015,huangNatComm2015}.  The results presented in this paper are obtained for the (CIF) lattice parameters \cite{MURRAYJLCM1976} $a = b = 3.4368$ \AA, $c=11.6442$ \AA. Both Ta and As are at 4a Wyckoff positions (0,0,u) with u=0 for Ta and $u=0.4177$
for As.  We used a 8$\times$8$\times$8 k-mesh for the sampling of the Brillouin zone for both DFT based and beyond DFT calculations.

Our results for DFT Local Density Approximation (LDA) and DFT generalized gradient approximation (GGA) calculations,
including (or not) spin-orbit coupling (SOC), are fully consistent with the band structures available in the literature (when considering the same lattice parameters and the same level of DFT calculations) \cite{WengPRX2015,XuScience2015,LvNatPhys2015,LvPRX2015,yang_weyl_2015,SunPRB2015,huangNatComm2015}.

\begin{figure*} 
	\includegraphics[width=0.4\linewidth]{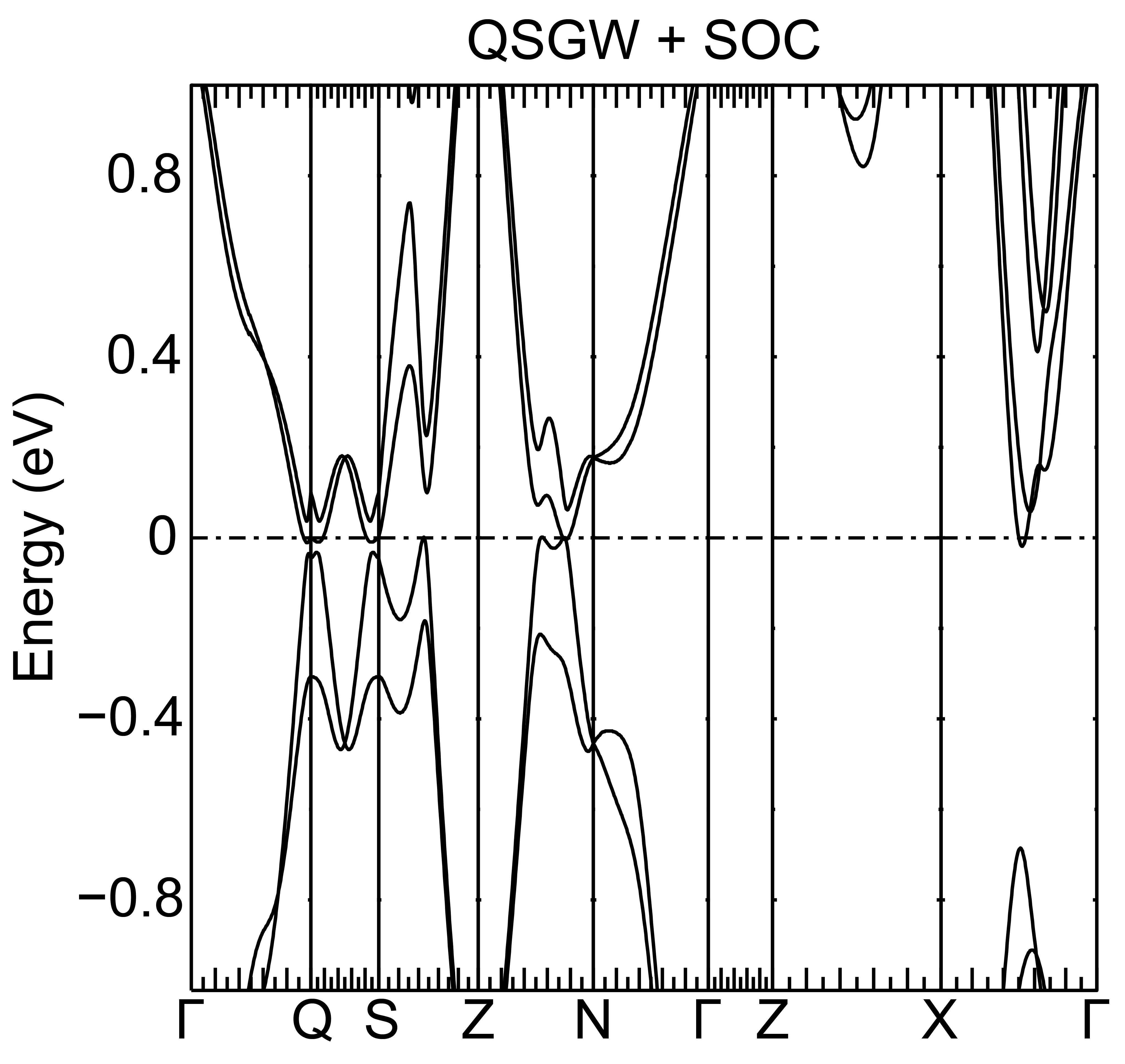}
	\caption{\label{fig:spaghetti} Calculated band structure of TaAs by QSGW+SOC.}
\end{figure*}

Going beyond DFT, we have also performed Quasiparticle Self-consistent \emph{GW} (QS\emph{GW}) calculations  shown in figure \ref{fig:spaghetti} \cite{Faleev04,mark06qsgw}.  For TaAs, the LDA and QS\emph{GW} are not so different, indicating that the system is weakly correlated.  Generally for weakly correlated systems, QS\emph{GW} is very high fidelity; thus we expect QS\emph{GW} results to yield an excellent description of the true one-particle Green's function. The QS\emph{GW} Fermi surface is very similar to the DFT one.  However, QS\emph{GW} dispersions are larger, which affects the position of the Fermi level when TaAs is doped.

QS\emph{GW} provides both a noninteracting Green's function \emph{G}\textsubscript{0} which is carried through to
self-consistency and can also be used to generate energy bands; and the interacting \emph{G} which contains a dynamical self-energy with lifetime effects. From \emph{G}, one can extract the corresponding spectral functions, which describes properly the excitations of the system, to be compared directly with ARPES measurements.

\begin{figure*} 
	\includegraphics[width=0.75\linewidth]{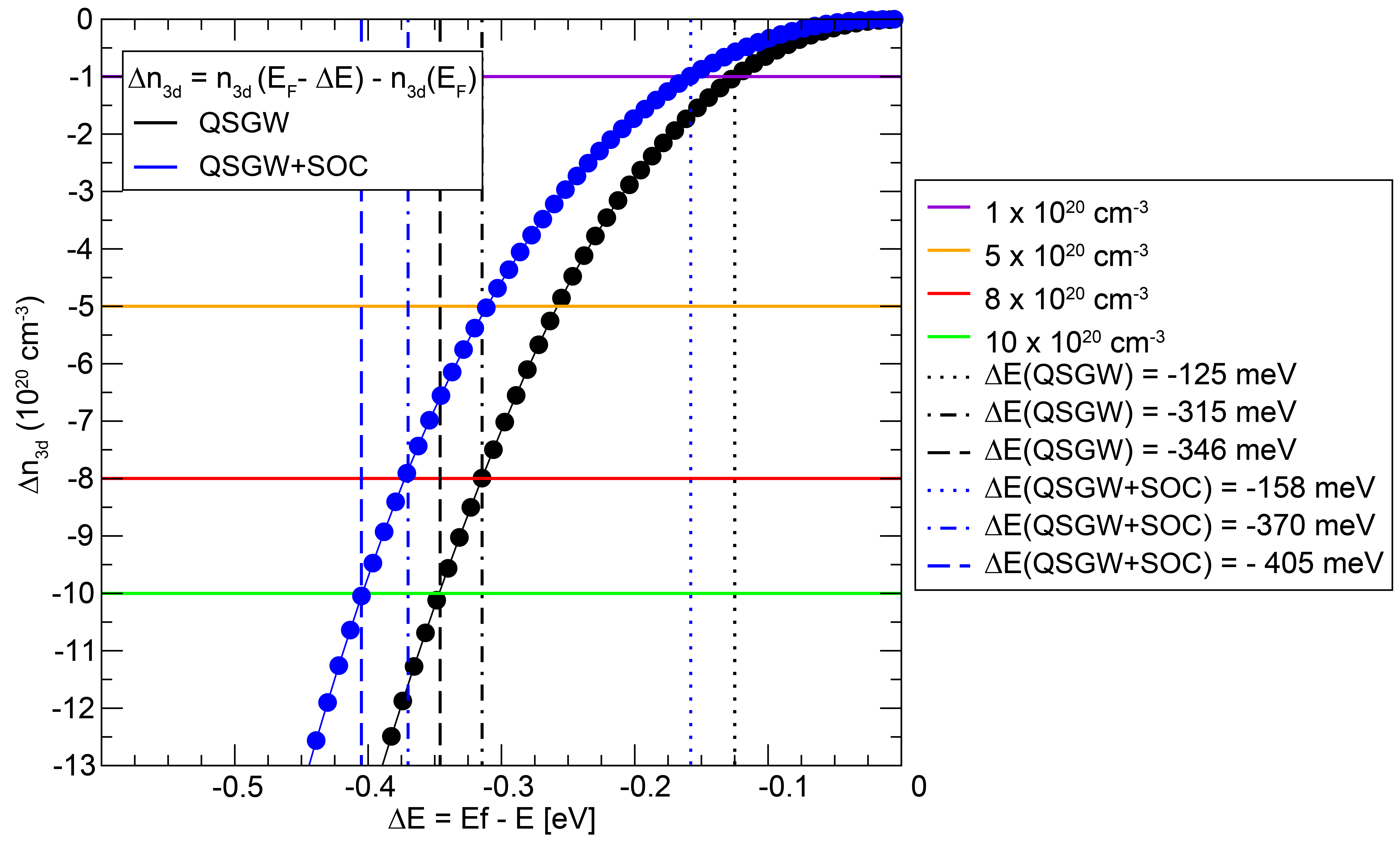}
	\caption{\label{fig:dNvdEf} a) The change in carrier concentration ($\Delta n_{3D}$) as a function of the location of the chemical potential. The $\Delta n_{3D}$ is relative to carrier concentration when the chemical potential is at the calculated Fermi level.}
\end{figure*}

\begin{figure*} 
	\includegraphics[width=1\linewidth]{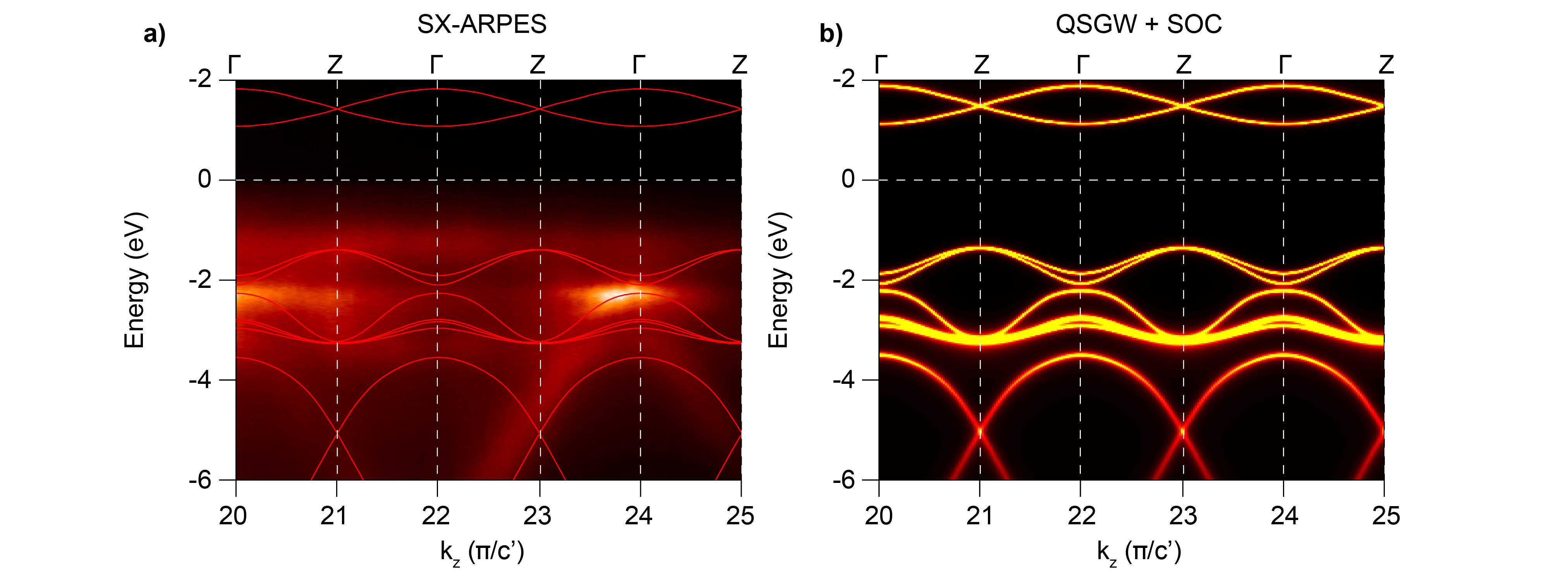}
	\caption{\label{fig:KZ} Dispersion of bulk states along $\Gamma-Z$: a) SX-ARPES k$_z$ scan calculated bands are overlaid an rigidly shifted by +300 meV. b) Calculated dispersion from the spectral functions. c' = c/2 as the k$_z$ dispersion follows a periodicity of half the c axis lattice parameter.}
\end{figure*}

In the main text we show that our films are unintentionally heavily hole doped due to defects which form during thin film synthesis. Thus to align with ARPES we make the usual assumption that bands shift rigidly with doping. We compute the Fermi level for a given hole concentration from the QS\emph{GW} Density-of-States (DOS). For hole-concentration of 1-10 $\times 10^{20}$ cm$^{-3}$, we can estimated a corresponding energy shift of the order of a few 100 meV as shown in Fig.~\ref{fig:dNvdEf}. The dispersion, along the k$_z$ direction, of the interacting band structure agrees well with the ARPES measurements, provided the +300 meV energy shift from doping is taken into account (Fig.\ref{fig:KZ}).

The two-dimensional constant energy contours and color intensity map obtained from the spectral functions at constant energy also agree favorably with the ARPES measurements (in the $k_x-k_y$ planes) when taking into account the Fermi energy shift corresponding to the hole-dopings.

\subsubsection*{Questaal's implementation}

Implementation of \emph{GW} requires both a 1-body framework and a two-body framework.  Both are described in detail in Questaal's methods paper, Ref.~\cite{Pashov:2020}, and the paper describing Questaal's implementation of QS\emph{GW} theory, Ref.~\cite{mark06qsgw}.

Questaal is an all-electron method, with an augmented wave basis consisting of partial waves inside augmentation spheres. The one-body basis consists of a linear combination of smooth, atom-centered Hankel functions as envelope functions, augmented by the partial waves.  Two partial waves are calculated at some linearization energy $\phi_\ell$ and energy derivative $\dot{\phi}_\ell$, which provides enough freedom to match the value and the slope to the envelope functions at the spheres' radius $r_\mathrm{MT}$.

\emph{One-particle basis}: In a conventional LMTO basis, envelope functions consist of ordinary Hankel functions, parameterized by energy $E$.  Questaal's smooth Hankel functions are composed of a convolution of Gaussian functions of smoothing radius $r_{s}$, and ordinary Hankel functions 
In the present work, $E$ is constrained to a fixed value ($-0.4$\ Ry), and $r_{s}$ determined by optimizing the total energy of the free-atom wave function.  These are kept fixed throughout the calculation, while the partial waves and linearization energy float as the potential evolves.  Another parameter is the sphere augmentation radius, $r_\mathrm{MT}$, which is chosen to make spheres touch but not overlap.  The result depends very weakly on the choice of $r_\mathrm{MT}$.

\emph{Two-particle basis}: The two-particle basis is needed to represent quantities such as the bare Coulomb interaction and the polarizability.  As with the one-particle basis, it has a mixed construction with interstitial parts and augmentation parts: envelope function products are represented as plane waves, since a product of plane
waves is another plane wave. Thus the interstitial parts of the mixed (product) basis are plane waves. 

\emph{Basis set and cutoff}: The Hamiltonian was constructed with $spdfgspdf$ orbitals (with envelope functions) on Ta,  and with $spdfspd$ orbitals on As. The core levels were integrated separately and treated at the Hartree Fock level. The Ta $5p, 4f$  and the As $3d$ partial waves were included in the Hamiltonian using local orbitals.
We used a cutoff of 4.4 Ry for the 2-particle basis, and of 5.4 Ry for the 1-particle basis.

\emph{k convergence}: The $GW$ mesh and the one-body mesh are generally different: the latter normally needs to be somewhat finer, as the self-energy is a relatively smooth function of $k$ while the kinetic energy is less so.  However, in the present 
work, we have used the same 8$\times$8$\times$8 mesh to make the \emph{GW} self-energy, for the one-body part.
	
\end{document}